\documentclass[journal=jacsat,manuscript=article]{achemso}

\usepackage[version=3]{mhchem} % Formula subscripts using \ce{}
\usepackage{adjustbox}
\usepackage{xcolor}
\usepackage{array}
\usepackage{multirow}
\usepackage{graphicx}
\usepackage{makecell}
\usepackage{cellspace}
\usepackage{booktabs}
\DeclareUnicodeCharacter{2212}{-}
\author{Janghoon Ock}
\affiliation[ChemE]
{Department of Chemical Engineering, Carnegie Mellon University, 5000 Forbes Street, Pittsburgh, USA}
\author{Chakradhar Guntuboina}
\affiliation[ECE]
{Department of Electrical and Computer Engineering, Carnegie Mellon University, 5000 Forbes Street, Pittsburgh, USA}
\author{Amir Barati Farimani}
\email{barati@cmu.edu}
\affiliation[MechE]
{Department of Mechanical Engineering, Carnegie Mellon University, 5000 Forbes Street, Pittsburgh, USA}
\title[An \textsf{achemso} demo]
  {Catalyst Property Prediction with CatBERTa: Unveiling Feature Exploration Strategies through Large Language Models}

\keywords{Computational Catalysis, Transformer, Large Language Model, Catalyst Screening \LaTeX}

\begin{document}

\begin{tocentry}

% Some journals require a graphical entry for the Table of Contents.
% This should be laid out ``print ready'' so that the sizing of the

% Inside the \texttt{tocentry} environment, the font used is Helvetica
% 8\,pt, as required by \emph{Journal of the American Chemical

% The surrounding frame is 9\,cm by 3.5\,cm, which is the maximum
% permitted for  \emph{Journal of the American Chemical Society}
% graphical table of content entries. The box will not resize if the
% content is too big: instead it will overflow the edge of the box.

% This box and the associated title will always be printed on a
% separate page at the end of the document.

\centering
\includegraphics[width=8.46cm]{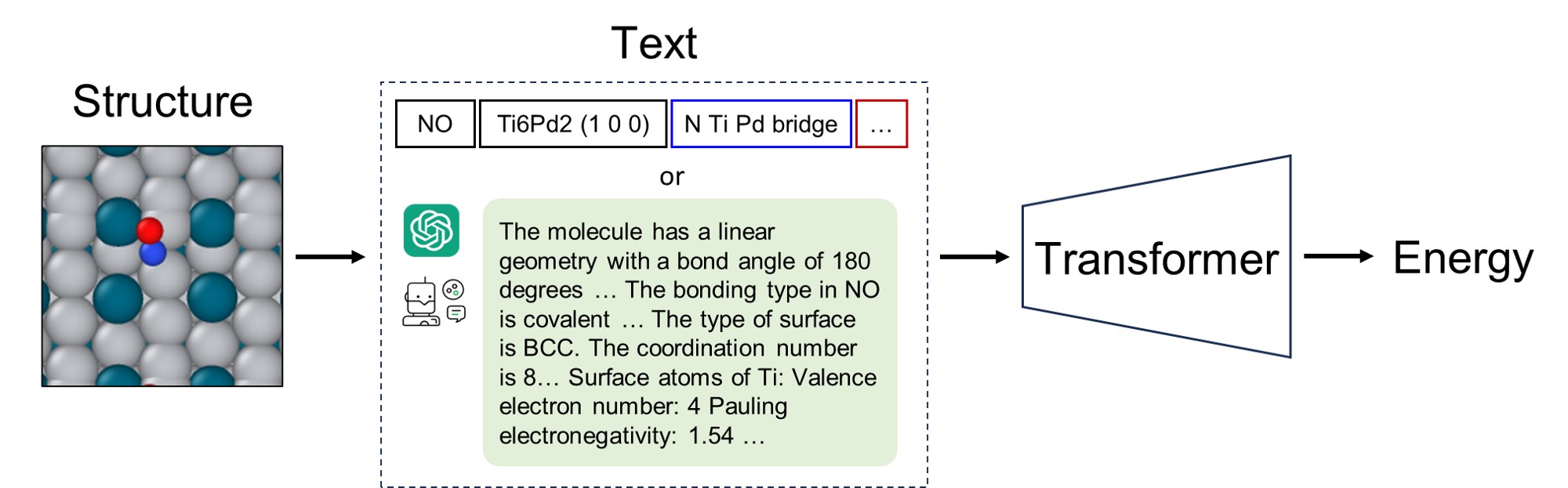} 
\label{fig:toc}

\end{tocentry}

\begin{abstract}

Efficient catalyst screening necessitates predictive models for adsorption energy, a key property of reactivity. However, prevailing methods, notably graph neural networks (GNNs), demand precise atomic coordinates for constructing graph representations, while integrating observable attributes remains challenging. This research introduces CatBERTa, an energy prediction Transformer model using textual inputs. Built on a pretrained Transformer encoder, CatBERTa processes human-interpretable text, incorporating target features. Attention score analysis reveals CatBERTa's focus on tokens related to adsorbates, bulk composition, and their interacting atoms. Moreover, interacting atoms emerge as effective descriptors for adsorption configurations, while factors such as bond length and atomic properties of these atoms offer limited predictive contributions. By predicting adsorption energy from the textual representation of initial structures, CatBERTa achieves a mean absolute error (MAE) of 0.75 eV—comparable to vanilla Graph Neural Networks (GNNs). Furthermore, the subtraction of the CatBERTa-predicted energies effectively cancels out their systematic errors by as much as 19.3\% for chemically similar systems, surpassing the error reduction observed in GNNs. This outcome highlights its potential to enhance the accuracy of energy difference predictions. This research establishes a fundamental framework for text-based catalyst property prediction, without relying on graph representations, while also unveiling intricate feature-property relationships.

\end{abstract}

\section{Introduction}
The search for optimal catalyst materials for specific reactions poses a significant challenge in the development of sustainable chemical processes. Traditional avenues of exploration have involved laborious experiments or computationally intensive quantum chemistry calculations, exemplified by Density Functional Theory (DFT) simulations \cite{CompCatal, Chen2021}. Nevertheless, the requirement to assess a vast array of systems makes the catalyst screening for optimality more challenging \cite{Brook2022, Tran2022}. This is attributed to the fact that a singular bulk catalyst can exhibit a range of surface orientations \cite{Brook2022, catalyst-surface, catalyst-surface2}. Additionally, adsorbates have the potential to bind to numerous distinct adsorption sites on these surfaces, with varying orientations \cite{adsorbml}. As such, relying solely on DFT calculations proves inadequate for swiftly assessing the vast array of potential adsorbate-catalyst combinations because of their time and resource demands. In response to these challenges, an increasingly prevalent approach involves harnessing the capabilities of machine learning (ML) methodologies to expedite the prediction of catalyst properties \cite{MLforCat, OC20_intro, OC20, Finetuna}.

In the field of molecular property prediction, Graph Neural Networks (GNNs) have emerged as a promising ML approach. They center on the graphical representation of molecular systems \cite{OC20, OC20_intro}, where atoms are depicted as nodes and bonds as edges, forming the primary input. Particularly in the context of catalysis research, these graph representations are generated by converting the structures of adsorbate-catalyst systems into graphs, effectively capturing the inherent structural intricacies of atomic arrangements. This enables the models to discern intricate connections between structures and properties \cite{cgcnn, schnet, dimenet, dimenetpp, gemnet-oc, scn}. However, the conversion of a 3D system structure into a graph mandates precise spatial coordinates for each atom, involving the meticulous identification of nearest neighbors within predefined proximity thresholds for individual atoms. The need for precise spatial comprehension, not easily attainable, can potentially introduce constraints during the initial phases of the screening process, especially when researchers aim to employ easily observable features for the screening procedure. Furthermore, GNNs function as a black box, making it difficult to assess the specific influence of distinct physical attributes within the system on property prediction.

The utilization of textual representations for describing adsorbate-catalyst systems presents an intriguing alternative to the graph-based approaches. Unlike graph representations, textual descriptions offer a natural way to incorporate observable features in a human-interpretable manner. Several established textual string-based representations exist for molecular and crystal structures, such as SMILES\cite{smiles}, SELFIES\cite{selfies}, InChI\cite{inchi}, and MOFid\cite{mofid}, each encoding the atomic system's structure in specific formats. Moreover, the emergence of generative language models has spurred ongoing research into generating textual descriptions for molecules, enhancing human interpretability \cite{christofidellis2023, molt5}. Once an atomic system is represented as textual data, it can be processed by deep neural networks. However, the segmentation of text into tokens presents a significant challenge, as placing essential tokens throughout the text's potentially extensive span can introduce complexities for various neural network architectures \cite{rajan2022}.

In recent years, an array of Transformer-based models, such as BERT \cite{devlin2019bert}, RoBERTa \cite{liu2019roberta}, GPT \cite{radford2018improving}, ELMo \cite{peters2018deep}, and LLAMA \cite{touvron2023llama}, have showcased exceptional proficiency across diverse natural language processing (NLP) tasks. The Transformer's distinctive strength lies in its adept application of an attention mechanism \cite{attn_all_you_need}, enabling the model to identify meaningful relationships among sentence tokens without relying solely on previous hidden states. This advancement has sparked a keen interest in exploring the Transformer's potential within chemistry and materials science \cite{chemberta, smiles_bert, stressd, transpolymer, moformer}. For instance, TransPolymer employs polymer sequence representations for predicting polymer properties, leveraging pretraining through masked language modeling (MLM) on unlabeled data \cite{transpolymer}. Similarly, MOFormer, a structure-agnostic Transformer model, utilizes text string representations of Metal-Organic Frameworks (MOFs) to predict their properties \cite{moformer}.

Drawing on the Transformer-based large language model's (LLM) advanced abilities in comprehending textual inputs, we propose the CatBERTa model, designed with the aim of predicting catalyst properties through textual representations. This distinctive trait opens a new avenue for property prediction in catalysis research, potentially bypassing the need for precise 3D atomic coordinates. Moreover, this approach enables us to exercise control over the inclusion of features in the input data in a way that is easily interpretable by humans. Scrutinizing self-attention scores provides insights into the varying importance levels assigned to different input features. Manipulating input data, coupled with interpretive analysis of self-attention scores, provides insight regarding the identification of the most effective features for predicting desired properties. Furthermore, CatBERTa serves as a litmus test for Transformer-based LLMs, assessing their capacity to grasp the scientific significance of specific features through data-driven methods.

The prediction of adsorption energy ($\Delta E_{\text{ads}}$) stands out as a primary focus in ML modeling for catalysis research. It serves as a pivotal descriptor directly linked to catalyst reactivity \cite{Norskov2009, Yang2014, BEP, BEP2, Ulissi2017}. The Open Catalyst 2020 (OC20) dataset encompasses an extensive collection of over 1.2 million DFT relaxations of adsorbate-catalyst systems \cite{OC20}. This dataset serves as the foundation for training ML models. Over successive development of GNNs for molecular property prediction have exhibited enhanced predictive accuracy for adsorption energies in the OC20 dataset. Illustrated by examples such as GemNet-OC \cite{gemnet-oc} and SCN \cite{scn}, leading-edge GNNs have achieved remarkable results in energy prediction. Notably, they have almost approached the DFT level accuracy by achieving mean absolute error (MAE) values as low as 0.28 eV \cite{ocp_lb}. Given the importance of adsorption energy prediction in catalysis research, our proposed CatBERTa model showcases its capability in property prediction for this task.

This research paper presents CatBERTa, a Transformer-based model designed to predict the adsorption energy of adsorbate-catalyst systems solely through textual inputs. CatBERTa improves interpretability by enabling researchers to integrate human-interpretable features into input data and identify the tokens in the textual representation that capture greater model attention. This understanding aids in pinpointing crucial catalyst attributes. Furthermore, the CatBERTa model demonstrates the efficacy and potential of Transformer-based LLMs for predicting catalyst properties. By integrating the capabilities of extensive language models with the demands of catalyst discovery, our goal is to streamline the process of effective catalyst screening.

\section{Results and discussion}

\subsection{CatBERTa framework}
\label{sec:framework}

\begin{figure*}[ht] %[h!]
\centering
\includegraphics[width=0.99\textwidth]{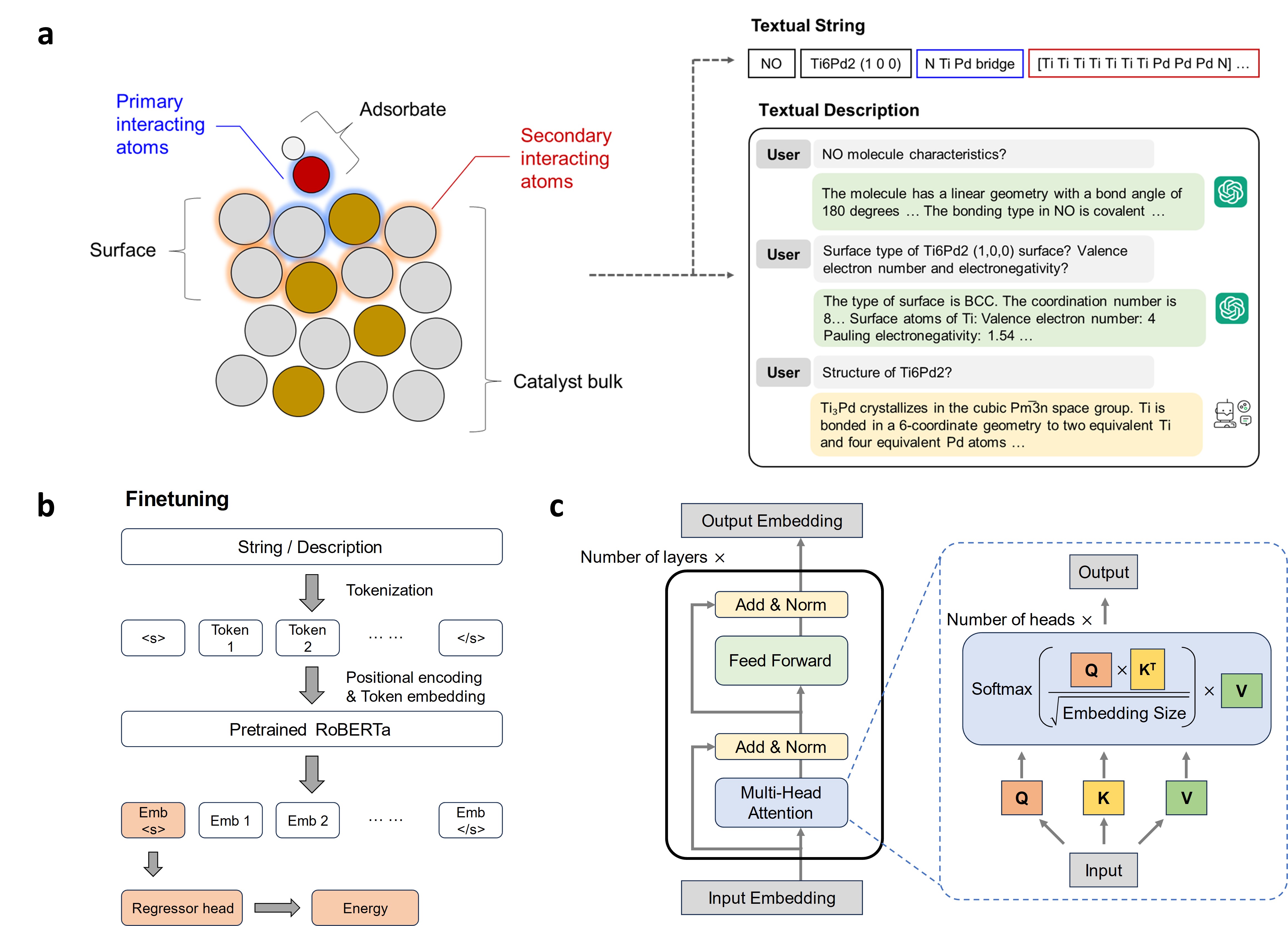} 
\caption{Overview of CatBERTa: \textbf{a} Transformation of structural data into textual format. The structural data undergoes conversion into two types of textual inputs: strings and descriptions. \textbf{b} Visualization of finetuning process. The embedding from the special token `\textless{}s\textgreater{}' is input to the regression head, comprising a linear layer and activation layer. \textbf{c} Illustration of the Transformer encoder and multi-head attention mechanism.}
\label{fig:framework}
\end{figure*}

We utilize a pretrained Transformer model, trained on a large corpus of natural language text, to predict the energy of the adsorbate-catalyst system, allowing us to thoroughly explore the feature space. To generate training data, we transform the initial 3D structures of DFT trajectories sourced from the OC20 dataset into easily understandable textual representations. Within this framework, we leverage publicly available datasets comprising 10k and 100k initial structures from DFT relaxations to create textual inputs for training the model. As illustrated in Figure \ref{fig:framework}a, our conversion process involves two distinct approaches: firstly, transforming the structure into a textual string containing exclusively the desired input features; secondly, generating a natural language description that elucidates the material's structure and characteristics.

At the core of our framework, we employ the encoder of RoBERTa, a robustly optimized variant of the Bidirectional Encoder Representations from Transformers (BERT) model. This serves as the engine for our framework. The generated textual inputs undergo tokenization and are then passed through the pretrained RoBERTa encoder (Figure \ref{fig:framework}b). Subsequently, the first token embedding from the encoder is used in the regression head to predict the scalar energy value. The model undergoes finetuning using both the 10k and 100k textual datasets, with detailed dataset information provided in the Methods section.

The foundation of Transformer encoders comprises a series of stacked self-attention and point-wise fully connected layers, as depicted in Figure \ref{fig:framework}c. In contrast to the architectures of recurrent neural networks (RNN), the Transformer model relies on the self-attention mechanism to establish meaningful connections between tokens positioned at different points within a sequence. This process of self-attention involves scaled dot-product attention, centered around the query, key, and value matrices. In our specific setup, the Transformer encoder consists of 12 hidden layers, each of which accommodates 12 attention heads. Additional details concerning the hyperparameters are available in Table S3.
%supplementary

\subsection{Input feature exploration}

The adsorbate-catalyst system involves the adsorption of an adsorbate molecule onto a catalytic surface. The amount of energy required for the adsorption hinges upon numerous factors such as the adsorbate and catalyst composition, surface orientation, and the particular adsorption configuration. Our research endeavors to unravel the underlying factors that drive energy predictions for adsorbate-catalyst systems, leveraging the Transformer which is pretrained for large language modeling purposes. In pursuit of our goal, we curate a diverse array of input text strings and descriptions sourced from the structural data in the OC20 dataset. Each of these inputs illuminates distinct facets of the adsorbate-catalyst system, contributing to a comprehensive understanding.

\begin{table}[htbp]
\centering
\caption{Different compositions of features in input strings and their corresponding mean absolute error (MAE) for the validation set. The outcomes result from training the model with a dataset of 100k, with the best performance case highlighted in bold text. Actual example strings are provided in Table S1.}
%supplementary
\label{tab:string}
\resizebox{\textwidth}{!}{
\begin{tabular}{clc}
\hline
No. & Textual String Format & MAE [eV] \\
\hline
String 1   & \textless{}s\textgreater{} adsorbate SMILES \textless{}/s\textgreater{} bulk composition (Miller index) \textless{}/s\textgreater{} & 0.85 \\
String 2   & String 1 + [primary interacting atoms, site type] & 0.79 \\
String 3   & String 1 + [primary interacting atoms, site type] [atomic properties] & 0.79 \\
String 4   & String 1 + [primary interacting atoms, site type] [secondary interacting atoms] & \textbf{0.75} \\
String 5   & String 1 + [primary interacting atoms, site type, bond distance] [secondary interacting atoms] & 0.77 \\
\hline
\end{tabular}
}
\end{table}

In crafting string-type inputs, we compose the textual string only with the essential target features in a condensed manner. This approach minimizes the inclusion of semantically vacant vocabulary, which could introduce unwanted noise. Our approach involves experimenting with five types of textual input strings. Initially, we concentrate on the information relevant to the adsorbate, bulk, and surface, deliberately excluding explicit adsorption configuration details. This initial approach results in an MAE of 0.85 eV. While this level of performance does not meet the stringent requirements of practical applications, it is crucial to acknowledge that even these preliminary features contribute to capturing the fundamental correlation (refer to Figure S1). 
%supplementary
Our subsequent efforts involve refining the input data by incorporating adsorption configurations. We begin by introducing information about the primary interacting atoms within both the adsorbate and the catalyst bulk. This leads to a decreased MAE of 0.79 eV, marking a 7\% reduction in comparison to the string type 1. Buoyed by this progress, we set out on a quest to enrich our feature set by introducing additional atomic properties, spanning atomic mass, periodicity, dipole polarizability, electronegativity, and electron affinity. Despite these enrichments, the resulting enhancements in accuracy are marginal. This prompts us to consider that attributes failing to enhance accuracy may already be inherently encoded within the model's training data through the inclusion of atom names.

We extend the adsorption configuration feature landscape by incorporating secondary interacting atoms within the catalyst bulk. The identification of atoms situated within a covalent radius of the primary interacting atoms on the surface is facilitated through the Pymatgen package \cite{pymatgen}. This refinement contributes to an improvement, reducing the MAE from 0.79 to 0.75 eV. This supports the great performance of GNN since the graph topology naturally embeds the primary and secondary interacting atoms by capturing the whole geometry. However, our attempt to further enhance accuracy by incorporating bond lengths of primary interacting atoms backfires, leading to an increase of MAE from 0.75 to 0.77 eV. This is due to the model overfitting when presented with distance information. As a result, we find that the distance feature does not function as a reliable generalizable predictor. In light of these findings, we can navigate the feature space, allowing us to easily evaluate the impact of adding specific target features.

\begin{table}[htbp]
\centering
\caption{Different compositions of features in input description and their corresponding mean absolute error (MAE) for the validation set. The outcomes arise from training the model using a dataset of 100k. If the model is exclusively trained using system descriptions, the resulting MAE is 0.79 eV.}
\label{tab:description}
\resizebox{\textwidth}{!}{
\begin{tabular}{c >{\centering\arraybackslash}m{2.5cm} m{7cm} c c}
\hline
\multicolumn{1}{c}{No.}  & Entity    & Input Feature                                                                                     & Tool          & MAE [eV]          \\ \hline
\multirow{3}{*}{Description 1} & System    & Adsorbate SMILES, bulk composition, Miller index, primary interacting atoms, adsorption site type & Pymatgen   & \multirow{3}{*}{0.84} \\ \cline{2-4}
                         & Adsorbate & Boding type, angle, length, molecule size, dipole moment, orbital characteristics                 & ChatGPT              &                       \\ \cline{2-4}
                         & Catalyst  & Composition, space group, bonding geometry, length, atom arrangement                                                             & RoboCrystallographer &                       \\ \hline
\multirow{3}{*}{Description 2} & System    & Adsorbate SMILES, bulk composition, Miller index, primary interacting atoms, adsorption site type & Pymatgen   & \multirow{3}{*}{0.79} \\ \cline{2-4}
                         & Adsorbate & Central atom, coordination number                                                                 & ChatGPT              &                       \\ \cline{2-4}
                         & Catalyst  & Surface type, coordination number, valence electron number, electronegativity             & ChatGPT              &                       \\ \hline
\end{tabular}
}
\end{table}

Unlike the previous string-based approach that requires manual feature integration, which can be a time-consuming task, we strive to streamline the process by automating the generation of input text. To achieve this, we leverage the capabilities of generative language models, specifically ChatGPT \cite{chatgpt3.5} and RoboCrystallographer \cite{robocrystal}, to generate textual descriptions in response to our queries. Embracing this approach empowers us to construct input text by incorporating targeted features. The resulting textual description comprises three primary sections: the initial part offers a comprehensive overview of the system, encompassing the adsorbate and catalyst bulk composition, surface orientation, and adsorption configuration; the subsequent segment delves into the adsorbate; and the last portion pertains to the catalyst. The query prompts employed to generate these descriptions can be found in Table S2.
%supplementary

To begin, we construct the text descriptions with general chemical and structural attributes, operating independently from prior knowledge of adsorption energy modeling. For an in-depth exploration of adsorbate characteristics, we extract details related to bonding types, molecular sizes, bond angles, lengths, and dipole moments through interactions with ChatGPT 3.5 version \cite{chatgpt3.5}. For bulk descriptions, we rely on RoboCrystallographer to generate textual representations of the catalyst bulk in the OC20 data. This specialized tool enables the generation of text-based depictions of crystal structures \cite{robocrystal}. This approach yields comprehensive textual depictions of the adsorbate and catalyst. When the model is trained solely on the system section, containing equivalent features to string type 2, the resulting MAE stands at 0.79 eV. This underscores the capacity of the language model to discern critical features through its inherent self-attention mechanism. However, the augmentation with adsorbate and catalyst bulk descriptions leads to a marginal increase in the MAE, reaching 0.84 eV. This rise indicates that the augmented features introduce some level of noise to the model without substantively enhancing data fitting.

In the subsequent refined phase, the description is enriched with scientifically grounded information, derived from the modeling study of adsorption energy \cite{Gao2020}. Recognizing the profound influence of factors like the coordination number of the central adsorbate atom and its interactions with counterparts on the surface, coupled with the electronegativity of these surface atoms \cite{Gao2020}, we engage ChatGPT-3.5 in generating textual portrayals of these features. This approach yields an MAE of 0.79 eV. Noteworthy is the fact that this MAE remains lower than the outcome derived from previous text descriptions encompassing general chemical characteristics. This discovery underscores that features within science-based descriptions do not compromise prediction accuracy, unlike the outcomes obtained from general chemical characteristics. Consequently, this revelation implies that coordination number and electronegativity hold a greater magnitude of significance compared to structural and characteristic descriptions of the adsorbate and bulk crystal. Our data-driven approach provides support to the modeling study conducted by Gao et al. (2020) \cite{Gao2020}, offering a cross-check methodology for the physics-based approach. This method enables the assessment of specific features' influence on property predictions.

\subsection{Self-attention visualization}

\begin{figure*}[hbtp] %[h!]
\centering
\includegraphics[width=0.99\textwidth]{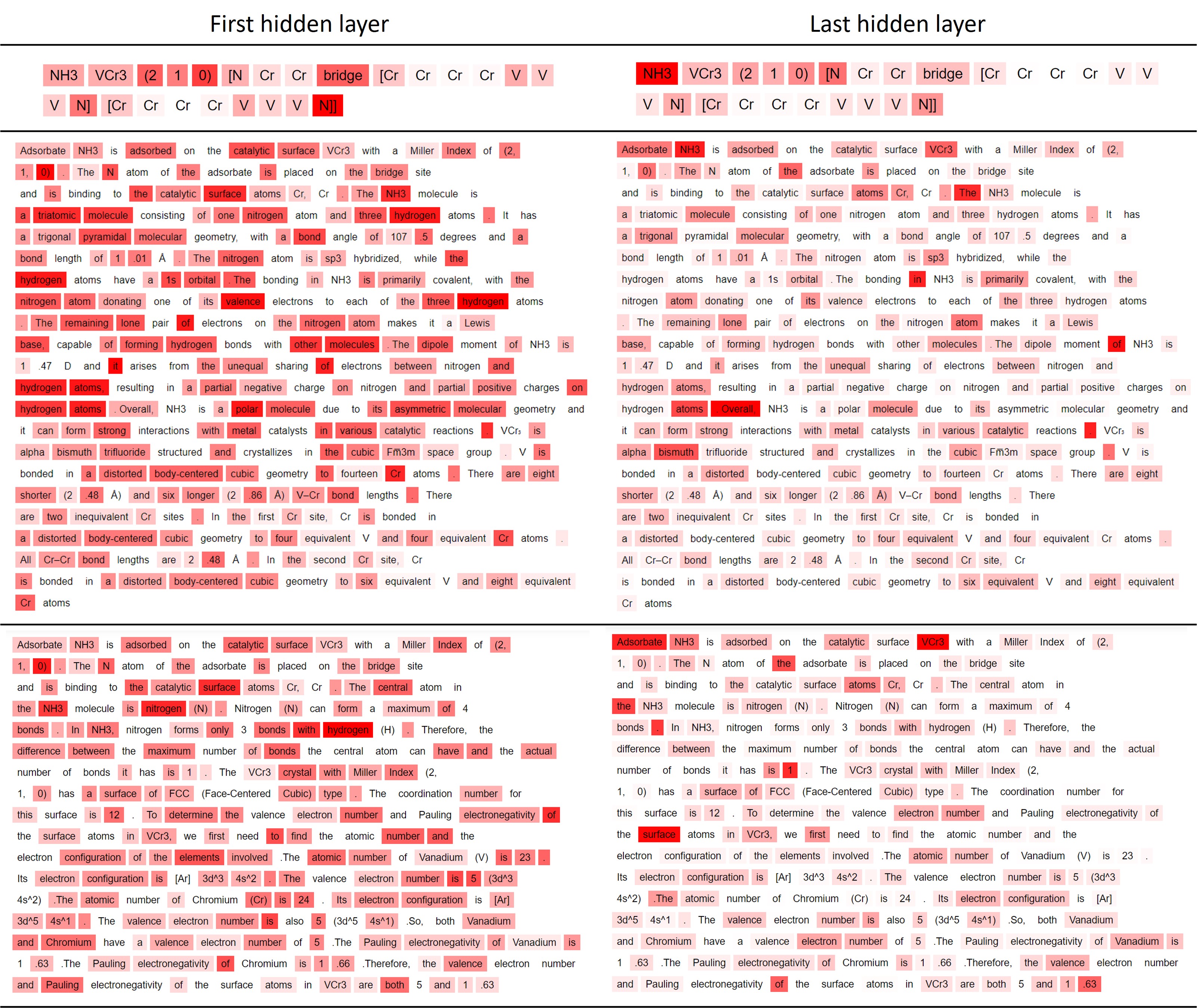} 
\caption{Visualization of attention scores from CatBERTa. The left column displays attention scores from the initial hidden layer, while the right column presents visualizations from the final hidden layer. The top row uses string type 4, the middle row has description type 1, and the bottom row displays description type 2 for the same system.}
\label{fig:attention}
\end{figure*}

The attention score, which acts as an indicator of the relationship between two tokens, offers insights into CatBERTa's acquisition of chemical knowledge and the individual contributions of each token to prediction outcomes. To illustrate, we choose an example adsorbate-catalyst system with \ce{NH3} as the adsorbate and \ce{VCr3} with a Miller index of (2,1,0) as the catalyst. This selection is based on systems displaying an MAE below 0.05 eV for all instances of string type 4 and description types 1 and 2. Employing the AttentionVisualizer package \cite{attn_visualizer}, we calculate the average attention scores from the 12 attention heads for each token and merge them for each vocabulary. In Figure \ref{fig:attention},  we present a visualization of these scores for both the initial and final hidden layers, where a stronger color represents a higher attention score.

In the attention score of the first hidden layer, strong relationships between nearby tokens are apparent, as evidenced by dispersed attention scores across the text. As the tokens pass through a sequence of hidden layers, they become concentrated on specific portions of the text. This transition leads to fewer vocabulary elements receiving high attention in the final hidden layer in comparison to the attention observed in the initial hidden layer.

When presented with input in string format, the model inherently places greater emphasis on the interactions between surface atoms and adsorbate atoms. Notably, in the context of secondary interacting atoms such as [Cr Cr Cr Cr V V V N] and [Cr Cr Cr Cr V V V N], where the initial Cr atoms within both brackets pertain to primary interacting atoms, the model exhibits heightened attention towards the N atom—an adsorbate atom engaged in interactions, despite the absence of explicit emphasis on the significance of interacting atoms in both entities.

Upon contrasting the attention scores from the final layer for text descriptions 1 and 2, a noticeable pattern emerges: the attention score in description 1 displays greater dispersion compared to that in description 2. The adsorbate and catalyst are the main emphasis of the model in the setting of description 2, with elements like bond count, electron amount, and electronegativity receiving secondary consideration. Conversely, in description 1, numerous vocabulary terms with limited semantic significance attract substantial attention. This could potentially contribute to the reduced accuracy observed in the case of description 1.

We can gain an initial understanding of interpretability regarding the model's emphasis on specific features. However, the attention scores might lack a comprehensive representation of the interaction between tokens and prediction outcomes \cite{Hao2021}. This disparity arises from the omission of Value matrix considerations in their evaluative approach. We acknowledge that a comprehensive resolution of the challenge in interpreting attention demands thorough additional investigation and concurrent research efforts.

\subsection{Features captured by latent space}

\begin{figure*}[htbp] %[h!]
\centering
\includegraphics[width=0.99\textwidth]{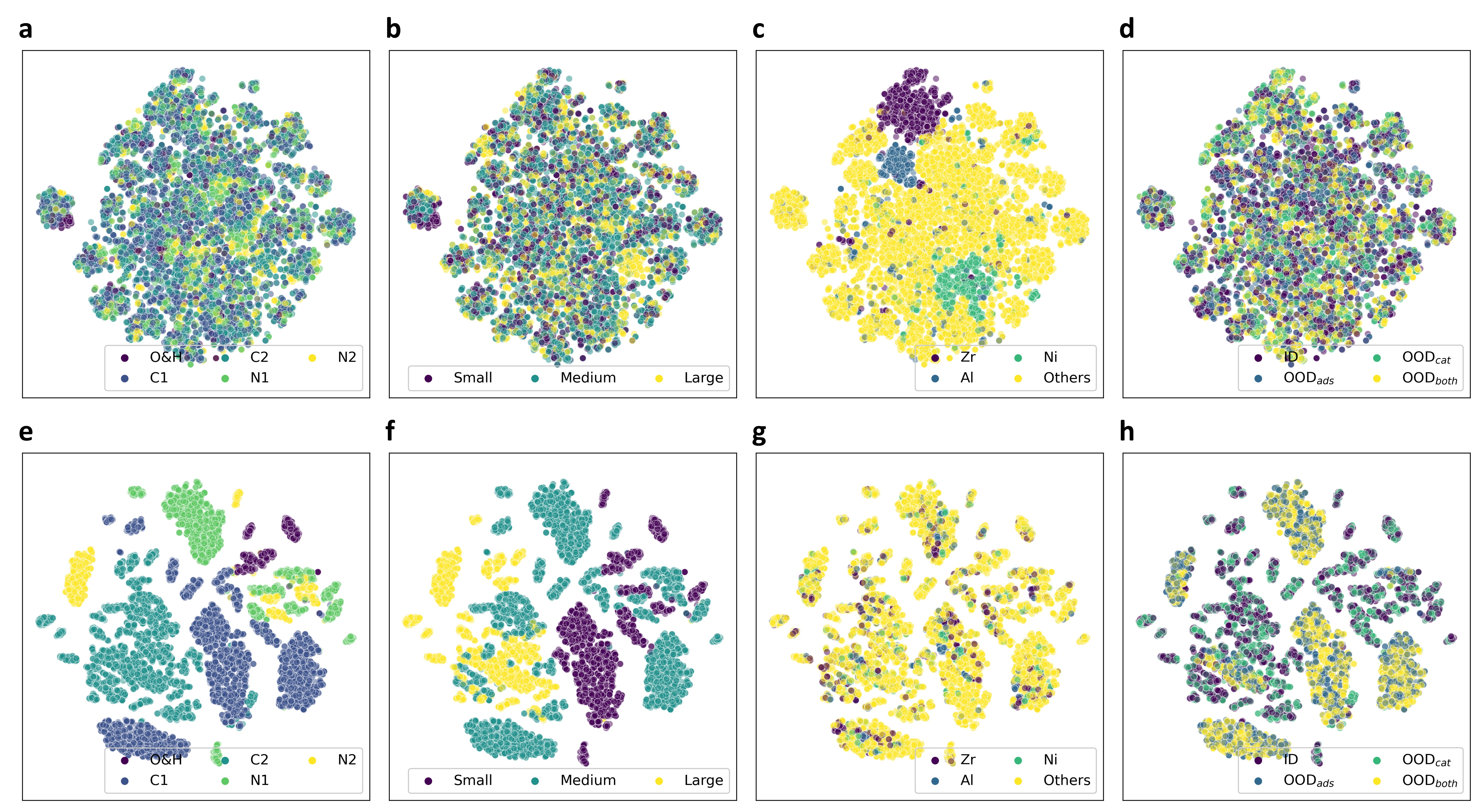} 
\caption{t-SNE visualization of first `\textless{}s\textgreater{}' token embeddings before and after finetuning. The upper row plots \textbf{a-d}  displays t-SNE plots before finetuning, while the lower row plots \textbf{e-h} represent after finetuning. \textbf{a} and \textbf{e} illustrate the adsorbate type. \textbf{b} and \textbf{f} indicate the size of the adsorbate molecule. For molecules with fewer than 3 atoms, it's small; for more than 5 atoms, it's large. \textbf{c} and \textbf{g} portray the bulk material which consists of certain atoms, specifically the top three common elements in the OC20 dataset: Zr, Al, and Ni. \textbf{d} and \textbf{h} depict the validation set splits. The in-domain split comprises adsorbates and catalysts seen during training, whereas the out-of-domain split includes adsorbates or bulk materials not encountered during training.}
\label{fig:tsne}
\end{figure*}

The impact of the training process can be illustrated by examining how the representations are captured in the latent space. This phenomenon is effectively visualized through the t-SNE plot shown in Figure \ref{fig:tsne}. The t-SNE plots specifically correspond to results from string type 4. The embeddings of the first `\textless{}s\textgreater{}' token, which traverse the regression head, are subjected to t-SNE analysis for visualization. The upper row of plots showcases t-SNE visualizations derived from a pretrained RoBERTa model that has not undergone our finetuning procedure with energy prediction. Conversely, the bottom row of plots corresponds to the t-SNE representations generated after our finetuning process.

Before undergoing finetuning, the embeddings exhibit a constraint in effectively capturing the type and size of the adsorbate. Embeddings associated with the same type and size of adsorbate do not cluster together; instead, they appear intermixed, as showcased in Figure \ref{fig:tsne}a-b. Conversely, the model demonstrates adeptness in distinguishing the bulk containing specific atoms. The embeddings stemming from systems containing \ce{Zr}, \ce{Ni}, and \ce{Al} form distinct clusters, as illustrated in Figure \ref{fig:tsne}c. Notably, given that a substantial proportion of atoms listed in both the primary and secondary interacting lists belong to the catalyst bulk, it is unsurprising that the model allocates more attention toward the bulk section in contrast to the adsorbate section.

In contrast, post-finetuning, the embeddings effectively capture distinctions in the type and size of the adsorbate. This is evident as embeddings featuring the same adsorbate type and size cluster together in the latent space, as depicted in Figure \ref{fig:tsne}e-f. Embeddings extracted from systems specifically containing certain atoms in the catalyst bulk no longer form distinct clusters but instead become intermixed, as shown in Figure \ref{fig:tsne}g. The finetuning process, conducted without explicit specifications of adsorbate types and sizes, directs the model's focus more keenly towards the adsorbate section.

Furthermore, notable is the model's competence in discerning the size of the adsorbate after finetuning, as this attribute can be deduced from the number of tokens allocated to the adsorbate section. Unlike the adsorbate type, which is inherently reflected in the tokens, the size relies on the quantity of tokens allocated to this section. Hence, this implies that the integration of positional embeddings enhances the extraction of more meaningful information from the provided tokens.

\subsection{Energy prediction results}

\begin{figure*}[htbp] %[h!]
\centering
\includegraphics[width=0.95\textwidth]{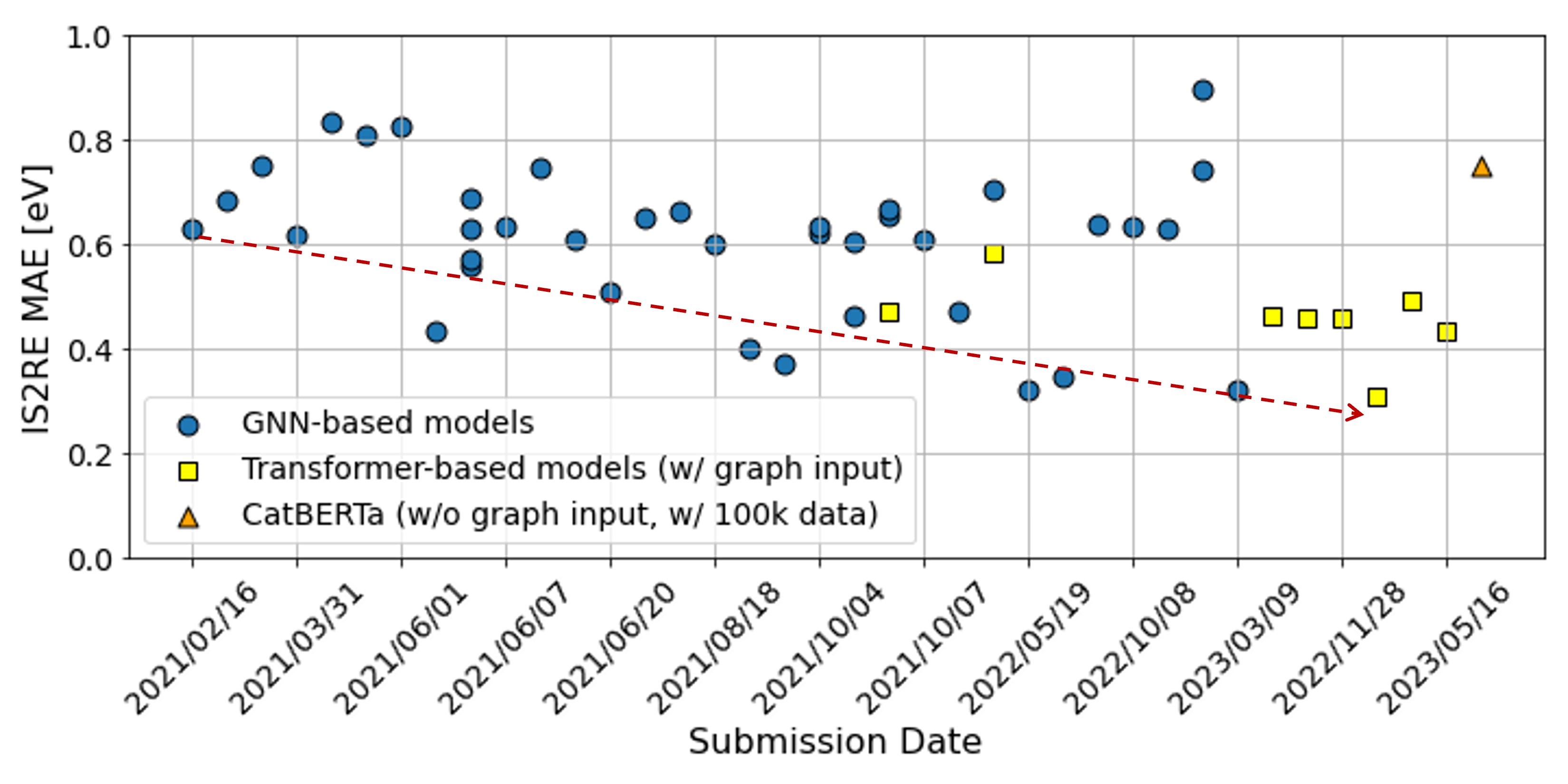} 
\caption{Open Catalyst Projct leaderboard results \cite{ocp_lb}. Model performance is evaluated using the initial structure to relaxed energy (IS2RE) task with the OC20 dataset.}
\label{fig:ocp_lb}
\end{figure*}

Since the project's inception, numerous models have been submitted, leading to a consistent reduction in overall MAE over time, as shown in Figure \ref{fig:ocp_lb}. GNNs have prominently driven this progress. Furthermore, the recent strides in submissions have been made by the emergence of Transformer-based models like Equiformer \cite{equiformer} and Moleformer \cite{moleformer}. These models have demonstrated noteworthy enhancements in performance, which closely parallel the ongoing surge in the prominence of the Transformer architecture. These state-of-art Transformer models require node and edge embeddings extracted from graph representations. However, our CatBERTa model, presented in this research, stands apart as the first Transformer-based approach that operates without the need for these graph embeddings. Instead, it leverages human-interpretable textual input to grasp a physical understanding of the system.

The prediction results of CatBERTa fall within the range of 0.3 to 0.9 eV, the results seen in the outcomes of the other submitted models. To evaluate the performance of CatBERTa, we conduct a comparative analysis against earlier versions of GNNs. Specifically, we employ CatBERTa trained and validated using string type 4 and compare its performance with publicly available checkpoints of CGCNN, SchNet, and DimeNet++, all of which were trained on the same dataset size as CatBERTa. Our evaluation encompasses distinct data splits, including in-domain (ID) samples drawn from the training distribution, as well as out-of-domain (OOD) samples for adsorbates (OOD\textsubscript{ads}), catalysts (OOD\textsubscript{cat}), and both entities (OOD\textsubscript{both}), which encompasses previously unseen adsorbate and catalyst compositions \cite{OC20}. 

When trained with the 10k dataset, CatBERTa demonstrates an overall MAE of 0.82 eV, surpassing the MAE values attained by both CGCNN and SchNet. This heightened performance becomes particularly evident in the in-domain and out-of-domain catalyst splits, which underscores the proficiency of CatBERTa model when trained on a limited dataset. However, its predictive capacity shows a relatively diminished performance when applied to unfamiliar adsorbate molecules.

Despite CatBERTa being trained on an expanded dataset of 100k instances, resulting in an appreciable 8.5\% reduction in MAE, this advancement remains comparatively modest when compared to the substantial improvements observed in other GNNs. These models exhibit MAE reductions ranging from 24\% to 32\%. For the 100k training case, CatBERTa falls short of surpassing its predecessors among GNNs in terms of performance. However, as depicted in Figure \ref{fig:ocp_lb}, the performance of CatBERTa aligns with the range of other submitted results. It is noteworthy that this level of performance is achieved using solely a 100k dataset, in contrast to the results achieved using the complete 460k training dataset.

\begin{table}[htbp]
\centering
\caption{Mean absolute error (MAE) values for energy prediction across various models.}
\label{tab:mae}
\begin{tabular}{llccccc}
\hline
 &  & \multicolumn{5}{c}{MAE [eV] (lower is better)} \\ \cmidrule(lr){3-7} 
Data Size & Model & Total & ID & OOD\textsubscript{ads} & OOD\textsubscript{cat} & OOD\textsubscript{both} \\
\hline
\multirow{4}{*}{10k} & CGCNN     & 0.83 & 0.88 & 0.85 & 0.80 & 0.79 \\
& SchNet    & 0.90 & 0.89 & 0.93 & 0.87 & 0.90 \\
& DimeNet++ & 0.76 & 0.77 & 0.80 & 0.72 & 0.75 \\
& CatBERTa   & 0.82$\pm$0.02 & 0.75$\pm$0.02 & 0.95$\pm$0.02 & 0.71$\pm$0.01 & 0.88$\pm$0.02 \\
\hline
\multirow{4}{*}{100k} & CGCNN     & 0.63 & 0.58 & 0.72 & 0.55 & 0.66 \\
& SchNet    & 0.61 & 0.61 & 0.61 & 0.64 & 0.58 \\
& DimeNet++ & 0.57 & 0.54 & 0.63 & 0.52 & 0.60 \\
& CatBERTa   & 0.75$\pm$0.01 & 0.65$\pm$0.01 & 0.90$\pm$0.01 & 0.61$\pm$0.01 & 0.86$\pm$0.02 \\
\hline
\end{tabular}
\end{table}

\begin{figure*}[htbp] %[h!]
\centering
\includegraphics[width=0.8\textwidth]{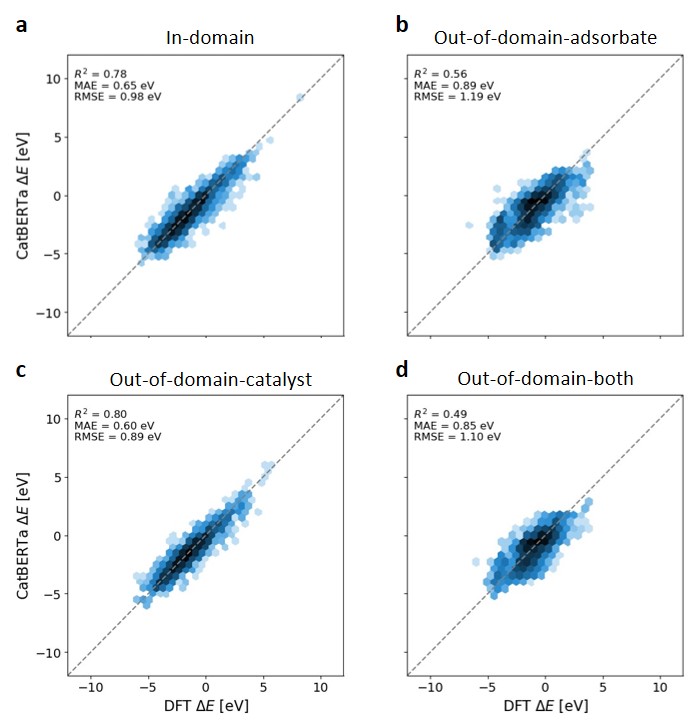} 
\caption{Parity plots for CatBERTa predictions. The x-axis corresponds to the label of DFT-calculated energy values, while the y-axis represents the energy values predicted by CatBERTa. \textbf{a} In-domain. \textbf{b} Out-of-domain-adsorbate. \textbf{c} Out-of-domain-catalyst. \textbf{d} Out-of-domain-both.}
\label{fig:parity}
\end{figure*}

\subsection{Error cancellation for chemically similar systems}

Every model shown in Figure \ref{fig:ocp_lb}, including CatBERTa, is designed to predict individual energy values. Nevertheless, the prediction of relative energy differences holds paramount practical implications compared to the individual energy prediction itself, notably in tasks such as screening for optimal catalysts employing volcano plots \cite{Li2019, Ooka2020, Cheng2008}, computing reaction energies \cite{Yang2014, Cheng2008, CO2RR}, and finding the most stable configuration \cite{Ulissi2017, adsorbml}. These applications hinge on discerning energy disparities between chemically similar systems. For instance, the energy difference between two adsorbed states can be derived by subtracting their respective adsorption energies. Similarly, a volcano plot can be constructed by contrasting the adsorption energies of adsorbate-catalyst systems featuring the same catalyst but varying adsorbates. Thus, an evaluation of energy difference prediction performance is imperative.

The difference in adsorption energy ($\Delta \Delta E_{ij}$) between two systems, denoted as $i$ and $j$, is computed by subtracting two machine-learning-predicted energies ($\Delta E_i$, $\Delta E_j$), as expressed in the equation below:

\begin{equation}
\label{eq:subtraction}
(\Delta E_i + \varepsilon_i) - (\Delta E_j + \varepsilon_j) = \Delta \Delta E_{ij} + \varepsilon_{ij}
\end{equation}

When energy differences are calculated by subtracting individual energy predictions, systematic errors are canceled out.  In cases where the signs of errors in these two energy predictions ($\varepsilon_i$, $\varepsilon_j$) align, there is the potential for a partial reduction in the error of the energy difference ($\varepsilon_{ij}$). This phenomenon is grounded in the principle of error propagation, elucidated in Supporting Information S5. Consequently, the precision of energy difference computation can be significantly enhanced if the errors associated with individual energy predictions are effectively canceled out through subtraction. Hence, the extent of error cancellation emerges as a significant metric, alongside the precision of individual energy prediction, capable of reflecting the potential to enhance the accuracy of energy difference calculations \cite{Ock2023}. 

Moreover, owing to the approximations inherent in DFT functionals that often give rise to systematic errors with correlations to the atomic structure of a system, these errors see considerable cancellation when comparing energies of akin systems \cite{Collins2020, Philipp2020, Hautier2012}. Consequently, machine learning models intended to surrogate DFT calculations should effectively replicate the error cancellation phenomena observed in DFT calculations.

The magnitude of error cancellation in ML models can be quantified with subgroup error cancellation ratio (SECR) as follows\cite{Ock2023}:

\begin{equation}
    \centering
    \text{SECR [\%]} = 1-\frac{\text{RMSE in subgroup}}{\text{RMSE in total pairs}}
    \label{eq:SECR}
\end{equation}

In this equation, the root mean square error (RMSE) denotes the standard deviation of the error distribution. The equation illustrates how much the error distribution narrows down within a specific subgroup of energy differences in comparison to the entire set of energy difference pairs. As a result, a higher SECR value indicates strong error cancellation, a factor that can enhance the precision of energy difference prediction. In the denominator, the set of total pairs represents all pairs in each split, including in-domain, out-of-domain-adsorbate, out-of-domain-catalyst, and out-of-domain-both. On the other hand, the numerator's subgroup is composed of pairs of systems within each split that possess at least one shared element, whether it is an adsorbate molecule or a catalyst. For instance, \ce{NH}-\ce{Al20Rh8} and \ce{NH}-\ce{N2Ti4} systems share the same adsorbate molecule as \ce{NH}, while \ce{OCH3}-\ce{Sc3Al} and \ce{COCH2O}-\ce{Sc3Al} share the common catalyst.

An interesting observation is that CatBERTa exhibits more pronounced error cancellation in the out-of-domain-adsorbate and out-of-domain-both splits compared to other GNNs. This is evident through its higher SECR values, reaching up to 19.3\%, as illustrated in Table \ref{tab:deltaE}. Notably, even leading GNNs like GemNet-OC \cite{gemnet-oc} do not achieve such a level of robust cancellation \cite{Ock2023}. Previous study reveals that a high embedding similarity leads to robust error cancellation \cite{Ock2023}. This observation finds further support in the t-SNE plots shown in Figure \ref{fig:tsne}d and h, where the embeddings of out-of-domain-adsorbate and out-of-domain-both are tightly clustered after the finetuning process. Hence, CatBERTa shows strong error cancellation in both splits, even though its accuracy in predicting individual energy values for these divisions falls behind other models. Particularly for the out-of-domain-both, CatBERTa exhibits a higher MAE of 0.2-0.3 eV when compared to other models in individual energy prediction. However, the discrepancy reduces considerably to 0.08-0.10 eV when it comes to energy differences, primarily due to the significantly stronger error cancellation. This distinctly highlights CatBERTa's potential to enhance error cancellation capabilities, thereby contributing to the accurate prediction of energy differences, as compared to GNNs.

% The MAE of energy prediction in OOD\textsubscript{ads} for CatBERTa is 6.6~20.4\% higher than other GNNs, while MAE in OOD\textsubscript{both} for CatBERTa is even 30.3~48.3\% higher than other models' results. On the contrary, for energy difference, the discrepancy between CatBERTa and other models is reduced to 0.15  

\begin{table}[htbp]
\centering
\caption{Mean absolute error (MAE) and subgroup error cancellation ratio (SECR) values across various models.  These results are obtained from the model trained using a 100k dataset.}
\label{tab:deltaE}
\begin{tabular}{lcccccccc}
\hline
& \multicolumn{4}{c}{MAE [eV] (lower is better)} & \multicolumn{4}{c}{SECR [\%] (higher is better)} \\ 
\cmidrule(lr){2-5} 
\cmidrule(lr){6-9}
Model & ID & OOD\textsubscript{ads} & OOD\textsubscript{cat} & OOD\textsubscript{both} & ID & OOD\textsubscript{ads} & OOD\textsubscript{cat} & OOD\textsubscript{both} \\
\hline
CGCNN & 0.86 & 0.94 & 0.81 & 0.87 & 1.7 & 9.1 & 4.0 & 8.1 \\
SchNet & 0.90 & 0.88 & 0.85 & 0.86 & 2.5 & 4.0 & 1.9 & 3.8 \\
DimeNet++ & 0.81 & 0.86 & 0.78 & 0.85 & 2.0 & 5.2 & 2.2 & 3.4 \\
CatBERTa & 0.95 & 1.01 & 0.87 & 0.95 & 2.7 & 18.2 & 3.8 & 19.3 \\
\hline
\end{tabular}
\end{table}

\section{Conclusion}

The introduction of the CatBERTa model in this study presents a Transformer-based approach for predicting energy using textual input data. By harnessing the power of a Transformer encoder pretrained on extensive natural language data, this model seamlessly incorporates features into input data in a format that is easily interpretable by humans. CatBERTa serves as a compelling example for large language models, demonstrating its capacity to comprehend the textual representations of the adsorbate-catalyst system. 

The ablation study involving distinct feature-laden input texts illuminates the significance of interacting atoms as descriptors for comprehending adsorption configurations. Conversely, the intentional inclusion of atomic properties does not contribute to accuracy enhancement. Considering that the interacting atoms within the adsorbate attract increased attention, we can infer that the model effectively captures their crucial role in adsorption energy prediction. As such, the attention score analysis enables the evaluation of emphasized features within the input text, distinguishing CatBERTa from GNNs. Moreover, through the finetuning process, the model becomes adept at prioritizing the adsorbate-related aspects as opposed to those associated with the catalyst.
 
CatBERTa's predictive accuracy falls within the performance range exhibited by other models submitted to the Open Catalyst Project, particularly comparable to the results achieved by earlier versions of GNNs. Although CatBERTa's energy prediction accuracy does not match that of state-of-the-art GNNs, it demonstrates an exceptional ability to cancel errors when subtracting the energies of similar systems, surpassing the magnitude observed in GNNs. This robust error cancellation underscores CatBERTa's potential in enhancing the accuracy of energy difference prediction.

In summary, we establish a foundational framework for predicting catalyst properties based on textual data, which holds relevance in early-stage catalyst screening. This framework helps researchers to estimate the properties of adsorbate-catalyst systems using easily discernible attributes such as adsorbate composition, bulk composition, surface orientation, and adsorption site. This approach enables preliminary assessments despite inherent accuracy limitations. Furthermore, our framework aids in the identification of effective descriptors for predicting catalyst properties through a data-driven methodology. By seamlessly integrating target features into the input text and subsequently training the model, we assess their contributions to improved accuracy, indicative of their capacity to encapsulate essential system characteristics. This methodological approach facilitates a comprehensive evaluation of how effectively these features capture relationships pertinent to property prediction.

\section{Methods}
\label{sec:method}
\subsection{Open Catalyst 2020 dataset}
\label{sec:oc20}
In this study, we utilized the Open Catalyst 2020 (OC20) dataset, which is the most comprehensive and diverse dataset available for heterogeneous catalysts \cite{OC20}. The OC20 dataset comprises 1.2 million DFT relaxations, employing the revised Perdew-Burke-Emzerhof (RPBE) functional \cite{rpbe}. Each relaxation involved approximately 200 single-point calculations, resulting in a dataset containing approximately 250 million structures along with their corresponding energies. 

The OC20 dataset contains training datasets of varying sizes for the initial structure to relaxed energy (IS2RE) task: 10k, 100k, and 460k. We focus on utilizing the 10k and 100k datasets for model training. Since the DFT relaxation trajectories for the test dataset in the OC20 are not publicly accessible, our evaluation and comparison of model performance are centered on the validation dataset. As for the validation set, we randomly selected 10k data points from the openly accessible OC20 validation dataset, distributing them evenly across four distinct splits: in-domain, out-of-domain-adsorbate, out-of-domain-catalyst, and out-of-domain-both. Through deliberate consideration of these distinct partitions, we intend to establish a balanced and representative validation set that effectively assesses the performance of the CatBERTa model.

\subsection{Transformer encoder}
\label{sec:transformer}

The Transformer encoder consists of multiple stacked layers, each containing two primary components: a multi-head self-attention mechanism and a position-wise feed-forward network. An essential feature of Transformers is their ability to process input tokens in parallel, unlike the sequential processing of RNNs.

The self-attention mechanism assigns varying attention scores to different tokens in the input based on their relevance to the current processing token. It operates using three vector representations: query (Q), key (K), and value (V). The attention scores are derived by calculating the dot product of the Q and K vectors, which are then passed through a softmax function. The result is used as a weight for the V vectors, aggregating information from different parts of the sequence. The ``multi-head" aspect signifies the use of multiple parallel attention layers, allowing the model to capture diverse types of relationships within the data.

Given that Transformers lack an inherent sense of order or position, positional encodings are added to the embeddings at the input layer to provide information about the position of each token in the sequence. These encodings ensure that the model can account for the sequence's order, an essential feature for tasks like translation or sequence prediction.

Each attention output is passed through a feed-forward network, identical across different positions but with different parameters for each layer of the encoder. It acts to transform the attention-derived features and is followed by layer normalization and residual connections, enhancing the model's training stability and convergence speed.

% In essence, the Transformer-based encoder leverages attention mechanisms and position-sensitive processing to offer a powerful, parallelized method of encoding sequential data, proving its efficacy across a broad spectrum of tasks in the machine learning domain.

\subsection{Pretraining and finetuning}

Pretraining, in the context of language models, involves training a model on an extensive corpus before finetuning it for a specific downstream task. In this study, we utilize the RoBERTa model \cite{liu2019roberta}, which underwent pretraining on an extensive collection of textual data sourced from the BookCorpus and English Wikipedia datasets, amounting to over 160GB. Unlike BERT \cite{devlin2019bert}, which masks 15\% of tokens in each sequence per epoch, RoBERTa employs dynamic masking, allowing masked tokens to change across epochs. This phase equips the model to predict masked words within sequences, discern syntactic and semantic structures, and absorb vast general knowledge from the training corpus. Given its established proficiency in interpreting English text, including atomic symbols, we opt to use the pretrained RoBERTa model without introducing any custom pretraining.

For the finetuning stage, the pretrained model's weights undergo slight adjustments using a more compact, task-specific dataset—specifically, 10k and 100k textual data drawn from OC20 dataset in this study. This step refines the model weights for the energy prediction task. In alignment with our regression objective, we substitute RoBERTa's classification head with a custom linear layer designed to produce a singular scalar value as output. The embedding of the first \textless{}s\textgreater{} token after passing through the encoder is fed into the regression head to generate an energy prediction.

We adopt the MAE as our chosen loss function, coupled with the AdamW optimizer. We employ a strategy called grouped layer-wise learning rate decay (gLLRD) \cite{llrd}, where the learning rate gradually decreases across groups of layers using a multiplicative decay rate. Group assignment is based on layer depth. Specifically, the lower four layers constitute group 1, with an initial learning rate set to $1 \times 10^{-6}$. The subsequent four middle layers form group 2, with their initial learning rate scaled by a factor of 1.75. The uppermost quartet of layers, located near the output, has their initial learning rate increased by a factor of 3.5. This strategic approach acknowledges that different layers assimilate distinct insights from sequences. Layers closer to the output focus on capturing local and specific information, necessitating higher learning rates. In contrast, lower layers near the input excel in understanding broader and more generalized knowledge. Moreover, we introduce early stopping mechanisms, which halt training when validation performance shows no improvement over a designated number of epochs, mitigating the risk of overfitting to the training dataset.

\begin{acknowledgement}

The authors acknowledge the Bradford and Diane Smith Fellowship for their pivotal funding support. Additionally, the authors express gratitude to Rishikesh Magar and Yayati Jadhav for enlightening discussions on Transformer-based models, and commend Brook Wander for her valuable guidance in analyzing the Open Catalyst 2020 dataset.

\end{acknowledgement}

\section*{Data Availability Statement}
Both the Python code and the data employed in this study are available on GitHub at the following link: \url{https://github.com/hoon-ock/CatBERTa}.

\begin{suppinfo}

% This will usually read something like: ``Experimental procedures and
% characterization data for all new compounds. The class will
% automatically add a sentence pointing to the information on-line:
The supplementary materials encompass the following elements: examples of textual inputs, interactions with ChatGPT, hyperparameters for CatBERTa finetuning, prediction results for each input case, principle of error propagation, data composition in energy difference calculation, and energy difference prediction results.

\end{suppinfo}

\bibliography{reference}

\end{document}

\tableofcontents
%\clearpage
% \appendix
% \counterwithin{figure}{section}

\newpage
\section{Textual data examples}

\begin{table}[htbp]
    \centering
    \normalsize
    \caption{Textual string and description example for NH$_3$-VCr$_3$ system from validation set}
    \label{tab:example}
    \footnotesize
    \begin{tabularx}{\linewidth}{c X}
        % \hhline{==}
        \hline
        Type & String / Description \\
        %\hhline{==}
        \hline
        String 1 & \textless{}s\textgreater{}NH3\textless{}/s\textgreater{}VCr3 (2 1 0)\textless{}/s\textgreater{} \\
        \hline
        String 2 & \textless{}s\textgreater{}NH3\textless{}/s\textgreater{}VCr3 (2 1 0)\textless{}/s\textgreater{}[N, Cr, Cr, bridge]\textless{}/s\textgreater{} \\
        \hline
        String 3 & \textless{}s\textgreater{}NH3\textless{}/s\textgreater{}VCr3 (2 1 0)\textless{}/s\textgreater{}[N, Cr, Cr, bridge]\textless{}/s\textgreater{}[H, 1, 1.01, 1, 4.51, 2.2, 0.75][N, 7, 14.01, 2, 7.6, 3.04, -1.4][Cr, 24, 52.0, 4, 78.4, 1.66, 0.67][V, 23, 50.94, 4, 97.34, 1.63, 0.52]\textless{}/s\textgreater{} \\
        \hline
        String 4 & \textless{}s\textgreater{}NH3\textless{}/s\textgreater{}VCr3 (2 1 0)\textless{}/s\textgreater{}[N Cr Cr bridge [Cr Cr Cr Cr V V V N] [Cr Cr Cr Cr V V V N]]\textless{}/s\textgreater{} \\
        \hline
        String 5 & \textless{}s\textgreater{}NH3\textless{}/s\textgreater{}VCr3 (2 1 0)\textless{}/s\textgreater{}[N (Cr 2.1) (Cr 2.1) bridge [Cr Cr Cr Cr V V V N] [Cr Cr Cr Cr V V V N]]\textless{}/s\textgreater{} \\
        \hline
        % \hhline{==}
        % Type & Description \\
        % \hhline{==}
        Description 1 & {\scriptsize Adsorbate NH3 is adsorbed on the catalytic surface VCr3 with a Miller Index of (2, 1, 0). The N atom of the adsorbate is placed on the bridge site and is binding to the catalytic surface atoms Cr, Cr. 
        
        The NH3 molecule is a triatomic molecule consisting of one nitrogen atom and three hydrogen atoms. It has a trigonal pyramidal molecular geometry, with a bond angle of 107.5 degrees and a bond length of 1.01 Å. The nitrogen atom is sp3 hybridized, while the hydrogen atoms have a 1s orbital. The bonding in NH3 is primarily covalent, with the nitrogen atom donating one of its valence electrons to each of the three hydrogen atoms. The remaining lone pair of electrons on the nitrogen atom makes it a Lewis base, capable of forming hydrogen bonds with other molecules. The dipole moment of NH3 is 1.47 D and it arises from the unequal sharing of electrons between nitrogen and hydrogen atoms, resulting in a partial negative charge on nitrogen and partial positive charges on hydrogen atoms. Overall, NH3 is a polar molecule due to its asymmetric molecular geometry and it can form strong interactions with metal catalysts in various catalytic reactions. 
        
        VCr$_{3}$ is alpha bismuth trifluoride structured and crystallizes in the cubic F$\overline{\text{m}}$3m space group. V is bonded in a distorted body-centered cubic geometry to fourteen Cr atoms. There are eight shorter (2.48 \r{A}) and six longer (2.86 \r{A}) V–Cr bond lengths. There are two inequivalent Cr sites. In the first Cr site, Cr is bonded in a distorted body-centered cubic geometry to four equivalent V and four equivalent Cr atoms. All Cr–Cr bond lengths are 2.48 \r{A}. In the second Cr site, Cr is bonded in a distorted body-centered cubic geometry to six equivalent V and eight equivalent Cr atoms.} \\
        \hline
        Description 2 & {\scriptsize Adsorbate NH3 is adsorbed on the catalytic surface VCr3 with a Miller Index of (2, 1, 0). The N atom of the adsorbate is placed on the bridge site and is binding to the catalytic surface atoms Cr, Cr. 
        
        The central atom in the NH3 molecule is nitrogen (N). Nitrogen (N) can form a maximum of 4 bonds. In NH3, nitrogen forms only 3 bonds with hydrogen (H). Therefore, the difference between the maximum number of bonds the central atom can have and the actual number of bonds it has is 1. 
        
        The VCr3 crystal with Miller Index (2, 1, 0) has a surface of FCC (Face-Centered Cubic) type. The coordination number for this surface is 12. To determine the valence electron number and Pauling electronegativity of the surface atoms in VCr3, we first need to find the atomic number and the electron configuration of the elements involved. The atomic number of Vanadium (V) is 23. Its electron configuration is [Ar] 3$d^3$ 4$s^2$. The valence electron number is 5 (3$d^3$ 4$s^2$). The atomic number of Chromium (Cr) is 24. Its electron configuration is [Ar] 3$d^5$ 4$s^1$. The valence electron number is also 5 (3$d^5$ 4$s^1$). So, both Vanadium and Chromium have a valence electron number of 5. The Pauling electronegativity of Vanadium is 1.63. The Pauling electronegativity of Chromium is 1.66. Therefore, the valence electron number and Pauling electronegativity of the surface atoms in VCr3 are both 5 and 1.63 respectively.} \\
        \hline
    \end{tabularx}
\end{table}

\newpage
\section{Interaction with ChatGPT}

% Description 1 and 2 are products of interactions with generative language models: ChatGPT and Robocrystallographer. In description type 1, the adsorbate section is formed by querying ChatGPT, while both the adsorbate and catalyst sections in description type 2 are generated through ChatGPT. Conversely, for the catalyst section of description type 1, information is extracted by querying the Materials Project ID to Robocrystallographer, as the catalyst bulk materials originate from the Materials Project database.

\begin{table}[ht]
\small
\setlength{\tabcolsep}{4pt}
\renewcommand{\arraystretch}{1.3}
\caption{Prompts for guiding ChatGPT-3.5 in generating descriptions. The terms `ads' and `bulk' correspond to the chemical symbols of the adsorbate and catalyst bulk, respectively.}
\label{tab:chatgpt}
\footnotesize
\begin{tabular}{|p{2cm}|p{0.7\textwidth}|}
\hline
Element & \multicolumn{1}{c|}{Query Prompt} \\
\hline
% & Description 1 \\
% \cline{2-2}
% & \multicolumn{1}{c|}{Description 1} \\
\hline
Adsorbate (type 1) & \{``role": ``system", ``content": ``You give correct and exact available scientific information about adsorbate molecules in the adsorbate-catalyst system, like bonding type, molecule size, bond angle, bond length, orbital characteristics, and dipole moment."\}, 

\{``role': ``user", ``content": f``Provide accurate and complete chemical information about the adsorbate \{ads\} molecule (or atom or radical). This should include details about the bonding type, molecular size, bond angles and lengths, orbital characteristics, and dipole moment, all of which must be valid and verifiable."\} \\
\hline
% & Description 2 \\
% \cline{2-2}
% & \multicolumn{1}{c|}{Description 2} \\
\hline
Adsorbate (type 2) & \{``role": ``system", ``content": ``The system will provide two pieces of information. Firstly, it will identify the central atom of the given molecule. Secondly, it will calculate the difference between the maximum number of bonds the central atom can have and the actual number of bonds it possesses."\}, 

\{``role": ``user", ``content": f``Provide me with the central atom in the \{ads\} molecule, and also the difference between the maximum number of bonds the central atom can have and the actual number of bonds it has. Consider that no hydrogen (H) atom is omitted in the provided SMILES string"\} \\
\hline
\multirow{2}{2cm}{Catalyst (type 2)} & \{``role": ``system", ``content": ``The system will provide the surface type, such as BCC (Body-Centered Cubic), FCC (Face-Centered Cubic), or HCP (Hexagonal Close-Packed), and the Coordination number based on the given bulk material and Miller index."\}, 

\{``role": ``user", ``content": f``\{bulk\} crystal with a Miller Index of \{miller\_index\}. Please provide me with the type of surface and its coordination number, keeping the answer simple."\} \\
\cline{2-2}
& \{``role": ``system", ``content": ``The system will provide valid valence and electronegativity values for the surface atoms in the bulk material provided by the user."\}, 

\{``role": ``user", ``content": f``Provide the valence electron number and Pauling electronegativity of the surface atoms in \{bulk\}. Assume that the surface is not oxidized and exclude surface effects. Keep the answer simple."\} \\
\hline
\end{tabular}
\end{table}

\newpage
\section{Hyperparameters for CatBERTa finetuning}
\label{si:hyperparameter}

\begin{table}[htbp]
	\centering
	\caption{Hyperparameters and architecture of CatBERTa}
	\label{tab:hyperparams}
	\begin{tabular}{lc}
	\hline
	Hyperparameter & Value \\
	\hline
	Max positional embeddings & 512 \\
	Number of attention heads & 12 \\
	Number of hidden layers & 12 \\
	Size of each hidden layer & 768 \\
	Batch size & 12 - 16 \\
	Optimizer & AdamW \\
	Initial learning rate & $1 \times 10^{-6}$ \\
	Early stopping threshold & 5 \\
	Warmup steps & 0 \\
	Loss function & MAE \\
	\hline
	\end{tabular}
\end{table}

\newpage
\section{Prediction results for each input case}

\begin{figure*}[htbp] %[h!]
\centering
\includegraphics[width=0.94\textwidth]{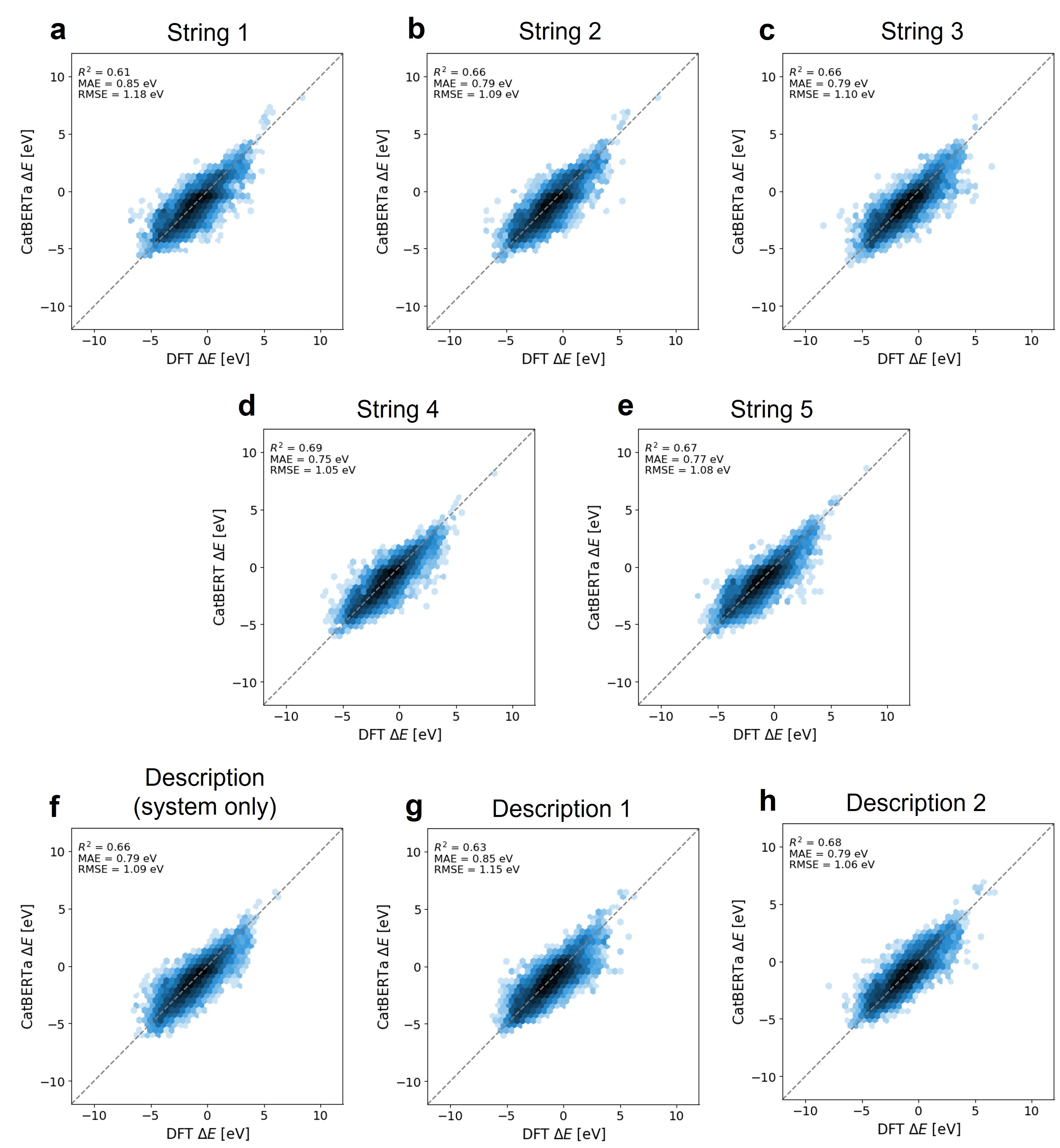} 
\caption{Parity plots of the validation set for each input case.}
\label{fig:input_parity}
\end{figure*}

\newpage
\section{Principle of error propagation}

When performing the subtraction of two values, the propagation of errors arising from these individual values can be described as follows:

% \begin{equation}
% \centering
% \varepsilon_{ij}^2 = \varepsilon_{i}^2 + \varepsilon_{j}^2 - 2 \sigma_{ij}
% \label{eqn:error_prob}
% \end{equation}

\begin{subequations}
\label{eq:error_prop}
\begin{gather}
(\Delta E_i + \varepsilon_i) - (\Delta E_j + \varepsilon_j) = \Delta \Delta E_{ij} + \varepsilon_{ij} \label{eq:subtraction}\\
\varepsilon_{ij}^2 = \varepsilon_{i}^2 + \varepsilon_{j}^2 - 2 \sigma_{ij} \label{eq:error}
\end{gather}
\end{subequations}

The error associated with the subtracted value ($\varepsilon_{ij}$) is influenced by the covariance ($\sigma_{ij}$) between the errors of the individual values ($\varepsilon_i, \varepsilon_j$). Thus, the cancellation of individual errors can occur during subtraction when two predictions are correlated and have positive covariance. However, if the predictions are independent or have negative covariance, the error in the difference will be larger than the individual errors.

This principle of error propagation finds direct relevance in our case. Here, we perform a subtraction of two machine-learning-predicted energies ($\Delta E_i$, $\Delta E_j$) to determine the energy difference ($\Delta \Delta E_{ij}$), as depicted in Equation \ref{eq:subtraction}. When the error signs of these two energy predictions coincide, there exists a possibility of partial error reduction. Conversely, if the errors exhibit opposing signs, it leads to error amplification, consequently magnifying the overall error in the subtracted outcome.

% \begin{equation}
% \centering
% (\Delta E_i + \varepsilon_i) - (\Delta E_j + \varepsilon_j) = \Delta \Delta E_{ij} + \varepsilon_{ij}
% \label{eqn:subtraction}
% \end{equation}

\newpage
\section{Data composition in energy difference calculation}

The validation set data is distributed almost evenly among four distinct categories: in-domain, out-of-domain-adsorbate, out-of-domain-catalyst, and out-of-domain-both. When calculating the energy difference, the process involves subtracting the two predicted values. As a result, the count of energy difference pairs aligns with the combination of the number of systems taken two at a time.

The assessment of chemical similarity among adsorbate-catalyst systems involves the evaluation of two key aspects: the resemblance found in the catalyst and the similarity observed in the adsorbate molecule. The column labeled `Sharing one element' indicates system pairs that share a single element—either the adsorbate or the catalyst. Conversely, the `Sharing two elements' column pertains to system pairs that share both the adsorbate and the catalyst. This indicates that the sole distinction between these two systems lies in their adsorption configurations. 

In the out-of-domain-adsorbate and out-of-domain-both splits, there exists a notable prevalence of system pairs that share both the adsorbate and the catalyst. This prevalence underscores a substantial similarity between these two systems. Consequently, this heightened similarity results in more pronounced error cancellation when compared to the other two splits, as shown in Table 4.

% \begin{table}[htbp]
% \centering
% \caption{Number, ratio, and sharing statistics for different systems and pairs.}
% \label{tab:sharing_stats}
% \begin{tabular}{lcccc}
% \hline
% & \multicolumn{2}{c}{Number {[}-{]}}  & \multicolumn{2}{c}{Ratio {[}\%{]}}  \\
% \cmidrule(lr){2-3}
% \cmidrule(lr){4-5}
% & \multicolumn{1}{c}{System} & \multicolumn{1}{c}{Pair} & \multicolumn{1}{c}{Sharing one element} & \multicolumn{1}{c}{Sharing two elements} \\
% \hline
% ID        & 2,493 & 3,106,278 & 0.02 & 1.78 \\
% OOD\textsubscript{ads}  & 2,494 & 3,108,771 & 0.02 & 17.34 \\
% OOD\textsubscript{cat}  & 2,507 & 3,141,271 & 0.16 & 1.80 \\
% OOD\textsubscript{both} & 2,506 & 3,138,765 & 0.10 & 20.05 \\
% \hline
% \end{tabular}
% \end{table}

\begin{table}[htbp]
\centering
\caption{Data composition and ratio of chemically similar pairs across each split.}
\label{tab:sharing_stats}
\begin{tabular}{lcccc}
\hline
& \multicolumn{2}{c}{Number {[}-{]}}  & \multicolumn{2}{c}{Ratio {[}\%{]}}  \\
\cmidrule(lr){2-3}
\cmidrule(lr){4-5}
Split & \multicolumn{1}{c}{System} & \multicolumn{1}{c}{Pair} & \multicolumn{1}{c}{Sharing} & \multicolumn{1}{c}{Sharing} \\
& & & \multicolumn{1}{c}{one element} & \multicolumn{1}{c}{two elements} \\
\hline
In-domain               & 2,493 & 3,106,278 & 0.02 & 1.78 \\
Out-of-domain-adsorbate & 2,494 & 3,108,771 & 0.02 & 17.34 \\
Out-of-domain-catalyst  & 2,507 & 3,141,271 & 0.16 & 1.80 \\
Out-of-domain-both      & 2,506 & 3,138,765 & 0.10 & 20.05 \\
\hline
\end{tabular}
\end{table}

\newpage
\section{Energy difference prediction results}

\begin{figure*}[htbp] %[h!]
\centering
\includegraphics[width=0.93\textwidth]{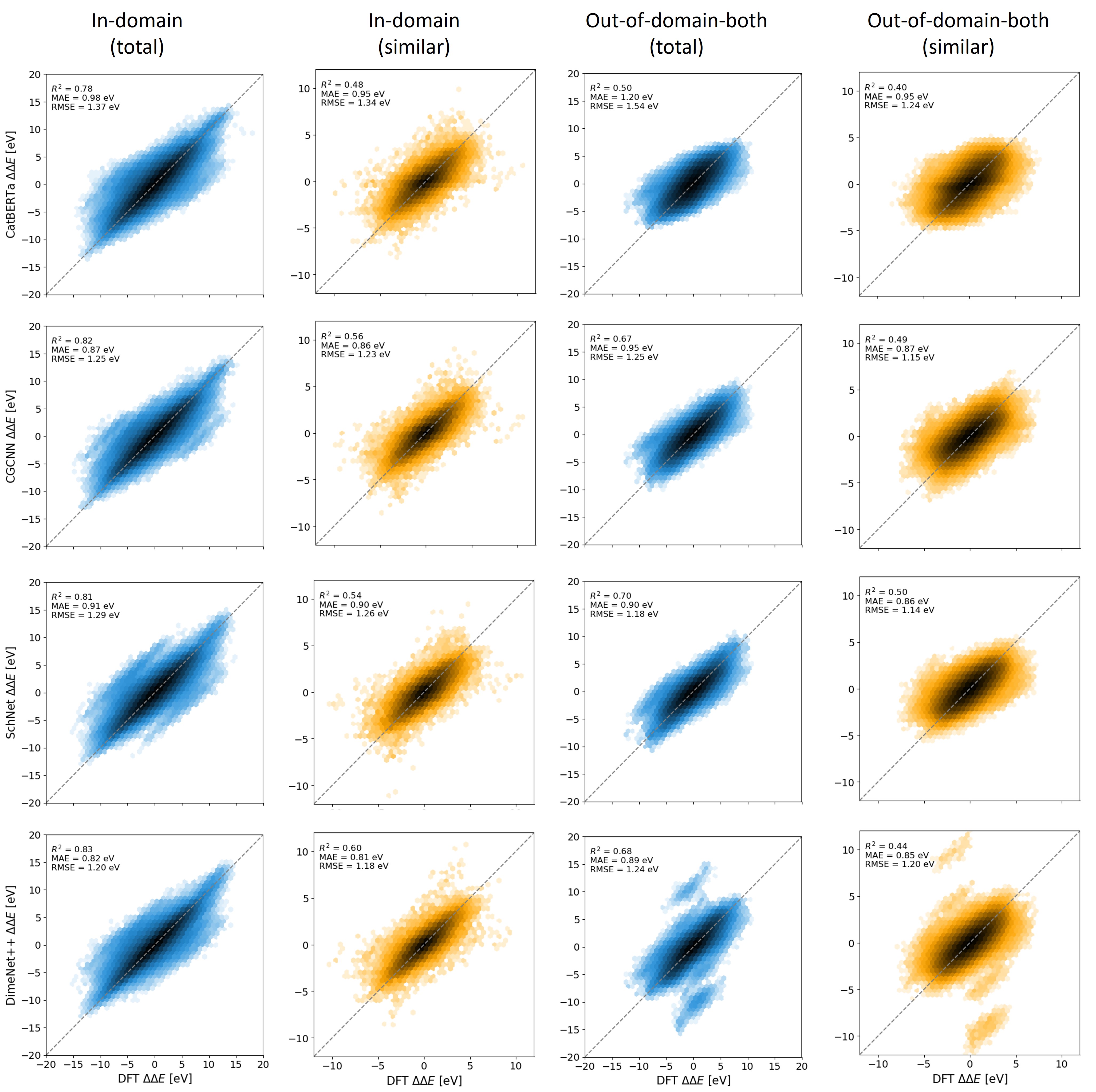} 
\caption{Parity plots of energy difference prediction. Columns 1 and 2 show results for all pairs and chemically similar pairs in the in-domain split, while columns 3 and 4 display outcomes for all pairs and chemically similar pairs in the out-of-domain-both split. Each row corresponds to a distinct model: CatBERTa, CGCNN, SchNet, and DimeNet++.}
\label{fig:en_diff}
\end{figure*}